# 2. The Chemical Composition of Mercury

Larry R. Nittler[1], Nancy L. Chabot[2], Timothy L. Grove[3], and Patrick N. Peplowski[2]

[1]Carnegie Institution of Washington, Department of Terrestrial Magnetism, Washington, DC 20015, USA.

[2]The Johns Hopkins University Applied Physics Laboratory, Laurel, MD 20723, USA.

[3]Massachusetts Institute of Technology, Department of Earth, Atmospheric, and Planetary Sciences, Cambridge, MA 02139, USA

## 2.1 Introduction

Chemical composition is a fundamental property of any planetary body. The bulk composition is determined by starting materials (the compositions of which are set by processes occurring in the Sun's protoplanetary disk) as well as secondary processes that may modify it, such as impacts with other bodies. Only materials at or near the surface of a planet, i.e., within its crust, are amenable to direct compositional measurements; interior (and hence bulk) compositional information must be inferred from indirect measurements and geophysical and/or geochemical considerations. The surface composition is influenced both by the bulk composition and geological processes such as planetary differentiation and volcanism as well as modification by impacts of other bodies. Prior to the MESSENGER mission, there were only a very few constraints on Mercury's elemental composition, namely the planet's anomalously high density, its surface reflectance properties, and the elements detected by ground-based astronomy in its weak exosphere.

Mercury's uncompressed (zero-pressure) density is 5.4 g/cm$^3$ (Anderson et al., 1987), much larger than those of the other terrestrial planets (e.g., 4.4 g/cm$^3$ for Earth and Venus). This high density clearly indicates a higher abundance of at least one heavy phase, of which metallic iron is by far the most plausible candidate given its high abundance in the bulk solar composition. Thus, it has long been recognized that Mercury must have a metallic iron-rich core that makes up a larger mass fraction than the cores of the other terrestrial planets (Siegfried and Solomon, 1974). The existence of a (partially molten) Fe-rich core was also suggested by the Mariner 10 discovery that Mercury has an internal magnetic field (Chapter 5), though an active dynamo was not required to explain the limited magnetic field data returned by that spacecraft.

Despite the implied high ratio of iron to silicate in the bulk composition, ground-based observations of Mercury have long suggested relatively low amounts of FeO in surface silicates. For example, near-infrared spectral absorption features associated with FeO in the silicate minerals olivine and pyroxene are commonly seen in lunar and asteroidal reflectance spectra but are weak or absent in Mercury spectra. Pre-MESSENGER spectral measurements (summarized in Chapter 8) generally indicated that surface silicates contained no more than ~3–5 wt% FeO, and possibly much less. Comparison of mercurian spectra with lunar spectra led some researchers to conclude that Mercury's surface may be rich in anorthositic rocks (dominated by the plagioclase mineral anorthite, $CaAl_2Si_2O_8$) like the lunar highlands and thus might have high Al and low Fe contents (McCord and Adams, 1972; Blewett et al., 1997).

Mariner 10 observations also revealed that Mercury is surrounded by a very thin atmosphere, a surface-based exosphere (Chapters 14 and 15); that spacecraft detected H, He and possibly O via ultraviolet spectroscopy. Ground-based observations later indicated the presence of Na, K, and Ca (Potter and Morgan, 1985; Potter and Morgan, 1986; Bida et al., 2000). Whereas the H and He most likely originate from the solar wind, the Na, K, and Ca were recognized as likely being derived from surface materials, removed via one or more processes (e.g., sputtering), thus indicating that these elements are present at Mercury's surface, although micrometeorite material might also contribute to the exosphere. However, uncertainties regarding the partitioning between exogenous and indigenous sources for exosphere constituents meant that these observations could not be used to infer surface composition information.

Armed with these compositional constraints as well as evolving theoretical concepts of planet formation, pre-MESSENGER researchers suggested a wide range of possible bulk and/or surface compositions for Mercury (Goettel, 1988; Taylor and Scott, 2003; Chapter 18). By and large, the most popular models to explain the planet's anomalously high density (e.g., giant impacts, evaporation by an early active Sun) invoked high temperatures, so it was broadly expected that Mercury would be a volatile-depleted world (the observations of volatile Na and K in the planet's exosphere notwithstanding). However, some volatile-rich model compositions were considered (e.g., Morgan and Anders, 1980; Goettel, 1988), as were models involving metal-rich and/or highly chemically reduced meteorites (Wasson, 1988; Burbine et al., 2002; Taylor and Scott, 2003). Wurz et al. (2010) derived estimates of Mercury's surface mineralogical and elemental composition from a combination of reflectance spectroscopic and exosphere observations.

The advent of the MESSENGER mission marked the possibility of testing these models with direct compositional measurements of Mercury's surface. The MESSENGER payload included three sensors for determining surface elemental composition (Section 2.2): the X-Ray Spectrometer (XRS), the Gamma-Ray Spectrometer (GRS), and the Neutron Spectrometer (NS). The GRS and NS were integrated into a single instrument (GRNS), but the sensors were independent in terms of placement on the spacecraft, functionality, and operation. During MESSENGER's flybys of Mercury in 2008–2009, compositional data over equatorial regions were obtained with both the GRS and NS. GRS observations yielded the unsurprising discovery that silicon was present on Mercury's surface as well as upper limits on the abundances of Fe, Ti, K, and Th (Rhodes et al., 2011). NS measurements indicated a surprisingly high level of neutron-absorbing elements (Lawrence et al., 2010) compared with expectations from modeled bulk compositions. By analogy with lunar neutron measurements and interpretation of MESSENGER Mercury flyby data (Denevi et al., 2009), the neutron data were interpreted as indicating relatively high surface abundances of Fe and Ti, postulated to be in the form of oxide minerals such as ilmenite ($FeTiO_3$). With the beginning of MESSENGER's orbital mission phase in March 2011, geochemical measurements of Mercury's surface with all three geochemical sensors began in earnest and quickly revealed the planet to have a volatile-rich surface with a surprisingly high abundance of sulfur and very low abundances of Fe and Ti (Nittler et al., 2011; Peplowski et al., 2011). This chapter summarizes the post-MESSENGER state of knowledge of Mercury's chemical composition, starting with the measured surface abundances of a wide range of elements and followed by estimates for the mantle and core compositions informed by the surface data and geochemical and petrologic considerations. We conclude by putting this information together to estimate the bulk composition of Mercury and place it into the context of planet formation models. A much more detailed discussion of the origin of Mercury, constrained in part by the compositional data discussed here, can be found in Chapter 18. Implications of the measured surface composition and its heterogeneity for the geological history of Mercury are discussed throughout the book, especially in Chapters 7, 8, 10, 11, 12, 18, and 19.

## 2.2 MESSENGER's geochemical sensors

MESSENGER's three geochemical sensors, the XRS, GRS, and NS, all exploited the interactions of high-energy radiation with the surface to remotely obtain information about surface composition through fundamental principles of atomic and nuclear physics.

The XRS relied on the method of remote-sensing X-ray fluorescence (XRF), whereby X-rays emitted by hot plasma and active regions low in the Sun's corona interact with atoms at an airless body's surface to induce emission of X-rays with energy characteristic of the fluorescing element. X-ray remote sensing is a surface technique, in that the detected X-rays come from within the top tens of micrometers of the planet's surface. Prior to MESSENGER, this method was successfully used to measure major-element surface compositions of the Moon (Adler et al., 1972a, 1972b; Narendranath et al., 2011; Weider et al., 2012a) and the asteroid 433 Eros (Trombka et al., 2000; Nittler et al., 2001). The XRS, described in detail by Schlemm et al. (2007) and Starr et al. (2016), consisted of three collimated gas-filled proportional counter detectors to detect X-rays from Mercury and a silicon-based detector pointed at the Sun to simultaneously measure the solar X-ray spectrum (critical for converting the planetary spectra into elemental abundances). A "balanced-filter" approach (Adler et al., 1972b) was used to deconvolve Mg, Al, and Si signals, as these X-ray lines are not resolved by the proportional counters (Fig. 2.1a). The 1–10 keV energy range of the instrument allowed measurement of the major and minor rock-forming elements Mg, Al, Si, S, Ca, Ti, Cr, Mn, and Fe. Because of MESSENGER's highly eccentric, near-polar orbit (Chapter 1), the spatial resolution of XRS measurements varied enormously both by latitude and time in the mission (due to the varying altitude of the spacecraft at periapsis), with measurement "footprint" sizes ranging from a few tens to hundreds of kilometers in the northern hemisphere and up to a few thousand kilometers over the south pole, observed when MESSENGER was at its farthest point in its orbit. Moreover, during typical (quiescent) solar conditions, the solar X-ray flux was sufficient to induce XRF from only Mg, Al, and Si; solar flares were required to enable measurements of heavier elements (Figure 2.1a). Since flares occur only sporadically and MESSENGER spent only a small fraction of each orbit over Mercury's northern hemisphere, the spatial coverage for the heavier elements in the northern hemisphere is incomplete. The methods used to analyze the XRS data and generate elemental abundance maps have been described in detail by Nittler et al. (2011) and Weider et al. (2012b, 2014, 2015).

<Insert Figure 2.1 near here>

The GRS detected gamma rays in the energy range 60 keV to 9 MeV, emitted both by the natural decay of radioactive isotopes and by nuclei through interactions with galactic cosmic rays (GCRs, i.e., high-energy particles, mostly protons, streaming through space and most likely originating in supernova explosions throughout the galaxy). GCRs interact with near-surface materials (to depths of a few meters), liberating neutrons from atomic nuclei. These neutrons subsequently interact with other nuclei, placing them in higher-energy excited states that subsequently decay via the emission of gamma rays at element-characteristic energies (Figure 2.1b). Compared with the shallow origin of planetary X-rays, however, the detected gamma rays arise from the top tens of centimeters of the surface. The GRS (Goldsten et al., 2007) consisted of a mechanically cooled high-purity crystal of germanium to detect gamma rays and measure their energy, surrounded on three sides by a plastic "anti-coincidence shield" (ACS). The ACS was used to discriminate background signals in the Ge detector from the gamma-ray signals of interest coming from Mercury. The mechanical cooler for the Ge crystal functioned nominally for about 14 months into the orbital mission before it failed, consistent with its expected lifetime, ending collection of gamma-ray data from Mercury early in MESSENGER's first extended mission (Chapter 1). Fortunately, a large amount of useful gamma-ray data was acquired prior to the failure of the cooler. GRS data were used to characterize C, O, Na, Al, Si, K, S, Cl, Ca, Fe, Th, and U concentrations on Mercury's surface; results and analysis procedures have been described in detail by Peplowski et al. (2011, 2012a, 2012b, 2014, 2015a) and Evans et al. (2012, 2015).

The NS characterized neutron emissions from Mercury in three broad energy bands: fast (~0.5 MeV to ~7 MeV), epithermal (~1 eV to ~500 keV), and thermal (~0.025 eV to ~1 eV). Fast neutrons are produced directly by GCR interactions with atomic nuclei, and the fast neutron flux measured by NS is proportional to the average atomic mass (<A>) of the regolith. Epithermal neutrons are the result of downscattering (slowing down) of fast neutrons via inelastic scattering, a process that is most efficient in the presence of H. As a consequence, the epithermal neutron flux is highly sensitive to hydrogen, and by inference water, content. Finally, thermal neutrons emissions are a balance between downscattering and neutron absorption, and the latter is sensitive to the bulk concentration of neutron-absorbing elements in the regolith. The NS consisted of three scintillator detectors: a central borated plastic detector that was sensitive to epithermal and fast neutrons and two Li-glass detectors that were sensitive to epithermal and thermal neutrons (Goldsten et al., 2007). MESSENGER NS data were used to confirm the presence of large amounts of water ice in permanently shadowed polar impact craters (Lawrence et al., 2013; Chapter 13). Thermal neutron measurements, made late in the orbital mission when the spacecraft flew at very low periapsis altitudes (Chapter 1), were used to infer the presence of carbon in Mercury's regolith (Section 2.3.3; Peplowski et al., 2016). Apart from H and C (see Section 2.3.3), MESSENGER neutron measurements were not used to investigate individual elemental abundances on Mercury, and we focus in this Chapter on XRS and GRS measurements of specific elements. However, neutron measurements acquired with both the NS and the GRS-ACS, which was re-purposed to provide measurements of neutrons complementary to those of the NS, have been used to derive important information on compositional variability across Mercury's surface, such as the definition of large-scale regions of distinct composition (Peplowski et al., 2015b; Lawrence et al., 2017; Chapter 7).

## 2.3 Mercury's surface composition

<Insert Figure 2.2 near here>

Mercury's average surface elemental composition, derived from XRS and GRNS measurements, is summarized in Table 2.1. Select maps of elemental abundances or abundance ratios are shown in Figure 2.2. GRS measurements of radioactive elements (i.e., K, Th, and U) are directly quantifiable, so for these elements absolute abundances are reported. For most of the other measured elements, abundances are reported as ratios to silicon. This choice reflects the fact that elemental ratios are more easily measured by the XRS and GRS methods and ratioing reduces or eliminates some systematic uncertainties from the data. Moreover, Si tends to vary less than other major elements among typical rock types, so variations in a ratio such as Mg/Si tend to reflect primarily variations in Mg rather than Si. GRS data support this statement, as Si gamma-ray emissions, and by extension Si concentrations, are observed to vary by just ~15% (standard deviation of the measurements) across the surface (Peplowski et al., 2012a). For those element ratios measured by both XRS and GRS (Al/Si, Ca/Si, S/Si, and Fe/Si), there is generally good agreement in the average values determined by the two techniques, providing confidence in the results and suggesting that Mercury's surface composition is similar at depths of micrometers and tens of centimeters. Where there are differences outside one-standard-deviation errors (e.g., Fe/Si), differences may reflect systematic errors in one or both of the techniques (e.g., Weider et al., 2014) or differences in sampling, due to depth or location. For example, GRS averages are over the northern hemisphere, whereas XRS maps for S/Si, Ca/Si, and Fe/Si are dominated by data in the southern hemisphere (Figure 2.2).

### 2.3.1 Volatile elements

MESSENGER's geochemical observations definitively refuted the notion that the surface of Mercury is depleted in volatiles relative to the other terrestrial planets. Highly volatile species, including H and organics, are present but limited to Mercury's permanently shadowed regions in polar impact craters (Chapter 13). Moderately volatile species, specifically Na, S, K, and Cl, are abundant and widespread (Nittler et al., 2011; Peplowski et al., 2011, 2012a, 2014; Weider et al., 2012b, 2015; Evans et al., 2015). The condensation temperatures (in Kelvin) for these elements (Lodders and Fegley, 1998) – 970 (Na), 674 (S), 1000 (K), and 863 (Cl) – are sufficiently low to provide a sensitive test of the maximum temperatures experienced by Mercury and/or its precursor materials during planetary formation processes. Mercury's surface has elemental concentrations of Na, S, K, and Cl (Table 2.1, Figures 2.2 and 2.3) that are similar to those found on the surface of Mars, long regarded as the most volatile-rich planet in the inner solar system. Mercury's Na, S, and Cl abundances are higher than those observed on the volatile-depleted Moon by an order of magnitude or more. Mercury's K concentrations are similar to those found within lunar regions rich in KREEP (K, rare-earth element, and P). However, lunar KREEP has elevated K levels as a result of the concentration of incompatible elements within the residues of fractional crystallization from a magma ocean, whereas on Mercury high K reflects a higher bulk volatile content, as discussed below.

<Insert Figure 2.3 near here>

Almost all elements detected on Mercury's surface exhibit clear spatial variability (Figure 2.2), and several "geochemical terranes" – regions with compositions distinct from their surroundings – have been defined (Peplowski et al., 2015b; Weider et al., 2015; Chapter 7). In principle, some of this heterogeneity for moderately volatile elements in Mercury's near-surface materials could arise from diffusion losses due to the extreme temperatures (up to ~700 K at places) experienced at the surface. Although the 50% condensation temperatures for the elements of interest (Na, S, K, Cl) are >670 K, diffusive loss can occur at substantially lower temperatures. For example, ~10 μm grains of Na-rich feldspars could lose sodium efficiently over a ~1 Gy timescale at temperatures of ~400 K (Kasper, 1975; Giletti and Shanahan, 1997), and potassium loss occurs at a temperature of ~475 K. Peplowski et al. (2012a) demonstrated an inverse correlation between K concentrations and maximum near-surface temperature for regions experiencing temperatures >350 K, and they suggested the possibility that K was being lost from sufficiently warm (equatorial) regions and being redistributed to cooler regions or lost to space. However, Weider et al. (2015) noted a similar inverse correlation between K and non-volatile Mg, and they argued that the correlation between K and temperature is coincidental on the grounds that Mg is not subject to thermal losses. Similarly, S concentrations are highly correlated with non-volatile Ca (Figure 2.4b), both being highest within Mercury's high-Mg region (Figure 2.2), an area at mid- to equatorial latitudes (Chapter 7). K, Na, and Cl all appear to have higher concentrations within Mercury's northern terrane (Peplowski et al., 2015), which is generally associated with the large smooth plains unit that dominates Mercury's northernmost latitudes (Head et al., 2011; Chapter 6). These observations argue against thermal control of the spatial distribution of moderately volatile elements on Mercury, and instead variability in K, Na, and Cl are attributed to differences in Mercury's geochemical terranes that result from having been derived from compositionally distinct magmas, suggesting that Mercury's mantle is compositionally heterogeneous or that the parent magmas underwent distinct igneous evolutions (Section 2.4, Chapter 7).

<Insert Figure 2.4 near here>

While direct measurements have revealed that Mercury's surface is not depleted in moderately volatile elements, the question of Mercury's bulk volatile content is more complex. The absolute abundances of elements over a planet's surface can vary appreciably as a result of differences in melt generation and crustal emplacement, along with subsequent modification. As a result, the high concentrations of volatile elements observed in near-surface materials are not necessarily indicative of Mercury's bulk composition. The ratio of the moderately volatile lithophile element K relative to the refractory incompatible element Th is commonly adopted as a proxy for the bulk volatile content of a planet, as K/Th is thought to be preserved throughout igneous processing and so the value at the surface should reflect the bulk value. MESSENGER revealed that Mercury's K/Th ratio is the highest observed in the inner solar system (Figure 2.3a), comparable to that of Mars and significantly higher than that for the volatile-depleted Moon (Peplowski et al., 2011, 2012a). This discovery argues against the simplistic expectation that volatile content in the solar system should vary inversely with solar distance (e.g., disk temperature). It also rules out many pre-MESSENGER theories for Mercury's formation, which predicted a near complete loss of volatiles from Mercury's surface and interior (Taylor and Scott, 2003; Chapter 18).

McCubbin et al. (2012) suggested that Mercury's high K/Th ratio may not be a reflection of its volatile content, but instead the result of sequestration of Th within Mercury's core. Specifically, they cited substantial uncertainties in elemental partitioning behavior under the highly reducing conditions inferred for Mercury (see Section 2.3.2). Although this issue is open,

Mercury's Cl content can provide additional insights into Mercury's bulk volatile content. Like K, Cl is a moderately volatile lithophile element (Lodders, 2003), and as a result the Cl/K ratio at the surface is expected to reflect the bulk value. That Cl is more volatile than K makes it a test of even lower maximum temperatures than are probed by the K/Th ratio. Mercury's Cl/K value is similar to those of both carbonaceous chondrite meteorites (thought to represent unfractionated solar composition) and Mars, yet significantly higher than those of Earth and the Moon (Figure 2.3b; Evans et al., 2015). The subchondritic Cl/K values observed on Earth and the Moon are attributed to volatile loss resulting from protracted impact and accretion histories (Sharp and Draper, 2013). In contrast, Mars is thought to have formed rapidly, with limited subsequent removal of material by impacts (Dauphas and Pourmand, 2011). Evans et al. (2015) argued that the similarity between the Cl/K ratios of Mars and Mercury suggests that Mercury might have also had a rapid accretion and limited impact history.

Mercury's high bulk volatile contents have had a direct influence on the geological history of the surface of the planet. For example, pyroclastic volcanism was widespread on Mercury, with numerous source vents observed across the surface (Robinson and Lucey, 1997; Head et al., 2008; Kerber et al., 2009; Chapter 11). Pyroclastic volcanism requires one or more magmatic volatiles, and combined spectral, XRS, and GRNS data strongly indicate that these volatiles were S- and C-bearing (Weider et al., 2016; Chapter 8 and 11). There is also strong evidence for flood volcanism on Mercury (Head et al., 2011). Smooth plains units are overwhelmingly located within Mercury's northern hemisphere (Chapter 6), and the largest unit, the northern smooth plains, is generally collocated with the highest K concentrations found on the planet. This scenario mimics the Moon, where volcanic plains units (e.g., lunar maria) are concentrated on the lunar near side and are generally associated with high KREEP content. Although a KREEP-like material has not been identified on Mercury, the concentration of incompatible elements (K, Cl) in the northern terrane raises the possibility that regions with concentrated abundances of heat-producing radioactive elements such as K may have been a factor in powering Mercury's flood volcanism. In addition to their influence on volcanism, volatiles appear to play an ongoing key role in Mercury's recent geologic evolution, in the formation of "hollows" (Blewett et al., 2011), shallow rimless depressions seen only on Mercury and discussed in detail in Chapter 12.

The pyroclastic deposits, some of which are as young as 1–3 Gyr (Goudge et al., 2014), and the even younger hollows provide evidence for recent and potentially ongoing release of volatiles from Mercury's interior. Relatively recently exposed volatiles would be subject to migration and cold trapping. Mercury's permanently shadowed regions (PSRs; see Chapter 13) are known to host extensive deposits of water ice and possibly C-bearing organic compounds (Lawrence et al., 2013; Neumann et al., 2013; Paige et al., 2013; Chabot et al., 2014a). Impact gardening is expected to obscure PSR-sequestered volatiles with a burial rate of 0.4 cm/Myr (Crider and Killen, 2005); therefore, the observation of numerous exposed water ice deposits with sharp and presumably young boundaries leads to the possibility of recent, and perhaps ongoing, delivery of volatiles to the PSRs. It is possible that the polar deposits include some fraction of material derived from volatiles outgassed from Mercury's interior.

### *2.3.2 Non-volatile elements*

XRS data acquired during the first few months of MESSENGER's orbital mission revealed Mercury's surface to have a major-element composition remarkably different from those of Earth and the Moon (Nittler et al., 2011). In addition to the surprising discovery that moderately volatile S is present at weight percent levels (Section 2.3.1, Table 2.1, Figure 2.2) compared with terrestrial crustal values of a few hundred parts per million (ppm), Mercury was found to have higher Mg/Si but lower Al/Si and Ca/Si ratios on average than the terrestrial or lunar crusts. Iron is present, but low, with an average abundance of ~1–2 wt% (Table 2.1) and a total observed range of Fe/Si of 0.02 to 0.1 (Weider et al., 2014). The relatively high Mg/Si and low Al/Si ratios of mercurian surface material (Figure 2.4a) rule out a plagioclase dominated crust as seen in the lunar highlands and previously suggested on the basis of spectral reflectance data (Section 2.1; Blewett et al., 1997). Although Mercury's surface composition was originally compared with that of terrestrial komatiites (Figure 2.4a; Nittler et al., 2011), due to the high Mg contents, subsequent petrologic modeling and experiments based on more complete elemental data sets indicated that the volcanic materials exposed on Mercury's surface are more similar to low-FeO terrestrial magnesian basalts and should be classified as norites (Stockstill-Cahill et al., 2012) or boninites (Vander Kaaden and McCubbin, 2016). As originally noted by Nittler et al. (2011), and confirmed by the full XRS data set for the entire mission, Ca/Si is highly correlated with S/Si (Figure 2.4b). Only the large pyroclastic deposit northeast of the Rachmaninoff impact basin has a composition clearly resolved from this correlation, and this departure is taken as evidence for loss of S-bearing volatiles during the explosive volcanism that generated the deposit (Section 2.3.1; Weider et al., 2016; Chapters 8 and 11).

Element maps (Figure 2.2) show that the surface of Mercury is remarkably heterogeneous chemically despite its relatively limited range of color and spectral properties (e.g., Chapter 8). Moreover, the geochemical terranes revealed by the elemental abundance maps are also seen, for example, in Mg/Si versus low-energy neutrons measured by the repurposed GRS-ACS (Section 2.2; Peplowski et al., 2015b) and in fast neutrons measured by the NS (Lawrence et al., 2017). The most chemically striking feature on the surface is a region of high Mg/Si, Ca/Si, S/Si, and Fe/Si, and low Al/Si situated at mid-northern to equatorial latitudes in the western hemisphere. In addition, the plains interior to Mercury's largest impact structure, the Caloris basin, are chemically distinct, with relatively high Al, but low Mg, Ca, S, Fe, and K abundances, compared with other smooth plains on the planet. The northern terrane, which includes much of the northern smooth plains (Head et al., 2011), is notably low in Mg, Al, Ca, and S, but enriched in the volatile elements K, Na, and Cl (Section 2.3.1). The presence of geochemical terranes on Mercury most likely points to melting of a heterogeneous mantle (Charlier et al., 2013; Weider et al., 2015; Namur et al., 2016a), as discussed in more detail in Chapter 7 (see also Section 2.4).

Nittler et al. (2011) noted that the major-element composition of Mercury's surface is similar, but not identical, to that found by partial melting of the highly reduced enstatite chondrite meteorites (McCoy et al., 1999; Burbine et al., 2002), and it is now recognized that the low iron and high sulfur contents of Mercury's surface are strong evidence of formation under highly chemically reducing conditions. There is decreasing incorporation of Fe and increasing incorporation of S into silicate melts as the availability of O goes down (Haughton et al., 1974; McCoy et al., 1999; Berthet et al., 2009; Namur et al., 2016b). The oxidation state of a magmatic system is often quantified in terms of oxygen fugacity ($f$O$_2$), which indicates the amount of O available to participate in chemical reactions and is commonly described on a logarithmic scale relative to an equilibrium reaction buffer. For example, the iron-wüstite buffer, denoted by IW, represents equilibrium in the reaction Fe + ½O$_2$ → FeO (see Section 18.2). McCubbin et al. (2012) and Zolotov et al. (2013) used the MESSENGER S and Fe data to estimate that Mercury's interior has $f$O$_2$ some 3 to 7 orders of magnitude below the iron-wüstite buffer (i.e. IW-3 to IW-7). More recently, Namur et al. (2016b) reported a large set of melting experiments under highly reducing conditions and through comparison with the MESSENGER data argued that Mercury's interior has log $fO_2$ = IW-5.4 ± 0.4. These results indicate that Mercury is more reduced than any other planet and most known solar system materials, with important implications for its formation (Chapter 18). Some of the consequences of Mercury's low oxygen fugacity for mantle and core compositions are discussed in Sections 2.4 and 2.5, respectively.

XRS measurements during solar flares also allow detection of X-ray fluorescence from the minor elements Ti, Mn, and Cr. All three are clearly present on Mercury's surface, but in low abundances (<1 wt%; Nittler et al., 2011; Weider et al., 2014; Murchie et al., 2015). Average abundance ratios relative to Si are provided for these three elements in Table 2.1; because of their low abundances, these elements have larger statistical errors and are more susceptible to systematic uncertainties than the major elements measured by the XRS. The low abundance of Ti, along with the low bulk Fe abundance (above), rules out the original interpretation of NS data acquired during MESSENGER's Mercury flybys (Section 2.1; Lawrence et al., 2010) as indicating substantial amounts of Fe-Ti oxides at the planet's surface. The high concentrations of thermal-neutron-absorbing elements observed in those measurements have subsequently been explained by the unexpectedly high Na and Cl content of Mercury's surface materials (Peplowski et al., 2015b).

The GRS has also provided measurements of the absolute concentrations of the naturally radioactive elements Th and U, which have mean northern hemisphere averages of 155±54 and 90±20 ppb, respectively. The decay of these elements, along with K (see Section 2.3.1), provides the primary source of long-lived heat generation for planetary interiors. Knowledge of K, Th, and U concentrations is therefore important for understanding the thermal and magmatic evolution of a planet. Directly relating the surface-measured concentrations to bulk values requires knowledge of the elemental partitioning behaviors of these elements, which are uncertain at the low $fO_2$ values predicted for Mercury. Nonetheless, the relative, time-varying concentrations of these elements can be characterized. This effort has shown that Mercury's initial heat production was higher than anticipated from pre-MESSENGER models of Mercury's composition, and that mantle heat production 4 Gy ago was a few times larger than it is today (e.g., Peplowski et al., 2011; Tosi et al., 2013). These observations are consistent with the evidence for widespread volcanic activity that largely ended shortly after the end of the late heavy bombardment (Chapters 11, 19). Mercury's Th/U ratio of about 2 is inconsistent with suggestions that U fractionated into Mercury's core (Malavergne et al., 2010). However it has been suggested that the GRS-measured U concentrations may be overestimated as a consequence of radon transport and diffusion within Mercury's regolith (Meslin and Déprez, 2012), as U gamma rays originate from a number of U-decay products, including radon.

### *2.3.3 Oxygen and carbon*

Oxygen itself is the most abundant element in most silicate rocks, typically making up 40–50 wt%, but it is often difficult to measure quantitatively. Thus, for laboratory geochemical measurements, it is in many cases determined indirectly by assigning a certain amount of O to each major cation given knowledge of or assumptions about the oxidation state. Thus, for example, since Mg has an oxidation state of +2, Mg is assumed to be present in the chemical form of MgO. Once the correct amount of O is assigned to every known element, the entire composition can be re-normalized to 100%, including the O, or differences from 100% can be used to infer analytical errors, the presence of other elements, or incorrect assumptions regarding oxidation state. This procedure has been used both as part of the XRS data analysis procedures (Nittler et al., 2011) and in estimating bulk compositions of Mercury for petrologic modeling and experiments (Stockstill-Cahill et al., 2012; Charlier et al., 2013; Vander Kaaden and McCubbin, 2016; Vander Kaaden et al., 2017). However, O produces characteristic gamma rays and GRS data can thus be used to infer the surface O abundance (or the O/Si ratio) directly. Evans et al. (2012) reported a GRS-derived O/Si weight ratio of 1.4, lower than inferred from typical stoichiometry for measured cations (e.g., ~1.7; Lawrence et al., 2013). However, as the GRS methodology for O was not well calibrated at the time, Evans et al. (2012) cautioned against over-interpretation of their reported value. Subsequent refinement of the methodology for estimating O/Si yielded an even lower value of 1.2±0.1 (McCubbin et al., 2017). Taken at face value, this anomalously low O/Si ratio would imply a very unusual surface mineralogy (Chapter 7). However, the low ratio may instead be due to secondary alteration of near-surface materials, for example by magmatic degassing in a process similar to industrial smelting (McCubbin et al., 2017), and hence may not be representative of the original composition of the lavas that formed the surface.

Mercury's surface is darker, on average, than that of the Moon (Robinson et al., 2008), and the darkest material on Mercury's surface (low-reflectance material or LRM) is associated with impact craters and their ejecta, suggesting that the reflectance-lowering component of Mercury's crust has an origin from some depth in the planet's crust (Denevi et al., 2009). On the Moon, low reflectance is largely driven by Fe- and Ti-bearing phases, but the discovery that surface abundances of Fe and Ti are low on Mercury (Section 2.3.2) ruled out the hypothesis that similar phases may be darkening that planet's surface. Murchie et al. (2015) carried out spectral modeling of MESSENGER reflectance data to show that C, in the form of fine-grained graphite, could match Mercury's global reflectance if present at ~1 wt% globally and ~5 wt% in the LRM. Peplowski et al. (2015a) developed a methodology to quantify C abundances from GRS spectra, and found a mean northern hemisphere average C concentration of 1.4±0.9 wt% (one-standard-deviation error). Adopting a three-standard-deviation threshold for positive identification led those authors to conclude that Mercury's surface has a mean C content of <4.1 wt%, consistent with the abundance needed to explain the planet's reflectance. The low altitudes reached by MESSENGER during its second extended mission allowed spatially resolved NS measurements of three deposits of LRM, all of which showed the neutron signature of enhanced C abundances (+1.1 to 3.1 wt%) relative to surrounding non-LRM materials (Peplowski et al., 2016), which have modeled C concentrations of ~1 wt%. These data, together with the spectral data of Murchie et al. (2015), convincingly demonstrate that weight-percent levels of C are present on Mercury. The association of higher C abundances with LRM, which is typically found within material excavated from depth by impact processes, indicates that the graphite is endogenic and may represent remnants of a primary graphite-rich crust formed by floatation in an early magma ocean and subsequently disrupted by impacts and buried by volcanism (Vander Kaaden and McCubbin, 2015) rather than solely by the addition to Mercury's surface of C from micrometeoriods (Bruck Syal et al., 2015). Moreover, although graphite itself is non-volatile, it is a potential source of volatile C-bearing compounds (e.g., through oxidation) that could contribute to pyroclastic volcanism and the formation of hollows (Section 2.3.1; Chapters 11, 12).

### 2.4 Mercury's mantle composition

As discussed in the previous section, the MESSENGER mission provided the first quantitative information on the chemical compositions and on the compositional variability of the extensive volcanic materials that cover the surface of Mercury. Mercury's surface lavas are unlike any others observed in the solar system. They are very low in Fe (and essentially FeO-free), sulfur-rich, and highly enriched in alkalis (Table 2.1). From crater size-frequency distributions it has been estimated that these volcanic deposits formed between 4.2 and 3.5 Ga (Marchi et al., 2013; Byrne et al., 2016; Chapter 6 and 9). The oldest volcanic rocks are found in the intercrater plains and heavily cratered terrain (IcP-HCT), and the youngest are found in smooth plains deposits, including the large expanse at high northern latitudes (Head et al., 2011; Chapter 6). A number of researchers have used petrologic modeling and experiments to infer candidate mineralogies and infer the melting conditions and mantle source compositions for the surface lavas (Stockstill-Cahill et al., 2012; Charlier et al., 2013; Namur et al., 2016a; Vander Kaaden and McCubbin, 2016; Vander Kaaden et al., 2017). One approach to this methodology, discussed in detail in Chapter 7, relates various geochemical terranes to the overall geological history of Mercury. In this section, we use the recent results of Namur et al. (2016a) to estimate the bulk composition of Mercury's mantle.

Namur et al. (2016a) performed high-temperature (1320–1580 °C), high-pressure (0.1–3 GPa) melting experiments under very reducing conditions (log $fO_2$ ~ IW-4 to -7; see Section 2.3.2) on synthetic analogs of the Mercury surface lavas. Compositions were chosen to correspond to the high-Mg portion of the ICP-HcT (Figure 2.2) and the low-Mg northern smooth plains (NSP). These compositions thus correspond to the "high Mg" and "northern terrane" compositions discussed in Chapter 7 and represent approximate end-members of Mercury's surface compositional range (Figure 2.4a). Namur et al. (2016a) mapped out a phase diagram for the selected compositions and identified multiple saturation points (points in pressure–temperature space at which two or more minerals are in equilibrium with the melt along the liquidus) involving forsterite + enstatite + liquid at 0.75 GPa and 1480 °C for the high-Mg IcP-HCT composition and at 0.75 GPa and 1380 °C for the low-Mg NSP composition. They also developed a thermodynamic model for the melting behavior of a lherzolitic source using experimental data from the system CMASN (CaO–MgO–$Al_2O_3$–$SiO_2$–$Na_2O$) and the MELTS/pMELTS algorithm (Ghiorso and Sack, 1995). The melting models predict an extent of melting ($F$ = melt fraction) of 0.46 and 0.27 for the high-Mg IcP-HCT and NSP lavas, respectively.

<Insert Figure 2.5 near here>

The low multiple saturation point pressures indicate shallow depths of melt segregation (~ 60 km) within Mercury. The most straightforward mechanism for producing these high extents of melting would be batch adiabatic decompression melting. By this mechanism, a parcel of mantle ascends in a mantle plume or as a mobile blob (Figure 2.5), it cools slightly as its temperature follows the adiabatic gradient (0.3 °/km), and it begins to melt at the depth where the adiabat intersects the solidus. The primary control on melt production is the temperature of the upwelling mantle; this effect on melting is illustrated in Fig. 2.5 for two different adiabatic paths. Magma is produced by the conversion of internal heat to heat of fusion. As a parcel of mantle continues to rise, melting continues and melt is extracted and transported upwards until a depth is reached at which the temperature of the parcel of mantle is lower than the solidus temperature for the residuum, i.e., the material remaining after melt is extracted. For Mercury lavas the melting rate is about 10 wt% per GPa change in pressure (or equivalent depth). For the high-Mg IcP-HCT lavas, melting would be predicted to begin at the base of Mercury's mantle near the core–mantle boundary at 400 km depth. For the NSP lavas, melting would commence at a depth of ~200 km. Therefore, the two end-member lavas sampled the mantle differently: one ascended from the base of the mantle, near the core-mantle boundary, and the second was the product of melting that started about half way between the core-mantle boundary and the surface. In both situations, on the basis of multiple-saturation-point pressures, the batch melts produced by decompression melting segregated at the same depth (0.75 GPa, or ~60 km). The compositions of the solids in the residuum assemblage are from the Namur et al. (2016a) experiments and are similar in composition (Table 2.2). The proportions of enstatite and forsterite from the experiments are also similar for both compositions: 45 wt% forsterite and 55 wt% enstatite.

With the phase proportions and compositions of enstatite and forsterite from the Namur et al. (2016a) experiments, the original composition of the mantle from which the two melts were produced can be estimated for each element $i$ from a mass balance equation:

$$W_i^{bulk} = F*W_i^{melt} + (1-F)*(\%Fo* W_i^{Fo} + (1-F)*\%En* W_i^{En})$$

where $W_i^c$ is the weight fraction of element $i$ in component c, $F$ = melt fraction, %Fo = 0.45, the mass fraction of forsterite, and %En = 0.55, the mass fraction of enstatite in the residue.

Estimates of the original composition of the mantle sources of the high-Mg IcP-HCT and NSP lavas obtained from this mass balance calculation are given in Table 2.2 for elements considered by Namur et al. (2016a). The mantle compositions inferred from the two magma types are similar in $SiO_2$, $Al_2O_3$, and MgO contents, but they differ in CaO and $Na_2O$, suggesting that the mantle of Mercury is heterogeneous with respect to these elements. However, the sulfur content of the estimated silicate portion of Mercury's mantle has not been included in these mantle compositions. The high inferred CaO concentration of the high-Mg IcP-HCT lavas may be a consequence of the high S content of these lavas (4 wt%) and the possibility that Ca may be in part partitioned into an oldhamite (CaS) component (Section 2.3.2, Figure 2.4b). On the basis of the solubility of S in reduced silicate melts, Namur et al. (2016b) concluded that at Mercury's oxygen fugacity condition of IW-5.4 ± 0.4, the mantle of Mercury contains 7–11 wt% S.

Many researchers have suggested that enstatite chondrites and/or bencubbinites (E and CB chondrites, respectively) could be analogs to Mercury's precursor materials (Wasson, 1988; McCoy et al., 1999; Taylor and Scott, 2003; Brown and Elkins-Tanton, 2009; Malavergne et al., 2010; Nittler et al., 2011). These meteorites are reduced metal-rich rocks that share many geochemical characteristics with Mercury, including high S, Na, and K, but very little FeO in the silicates. The silicate portions of two E chondrites and a CB chondrite (Jarosewich, 1990) are shown in Table 2.2. They are higher in $SiO_2$ and lower in MgO than the Mercury mantle estimates inferred from the surface lavas. However, Si becomes more soluble in metallic melts under highly reducing conditions (Berthet et al., 2009; Malavergne et al., 2010; Chabot et al., 2014b; Namur et al., 2016b) and thus may be an important component of Mercury's core (Section 2.5). We therefore also removed 20% of the $SiO_2$ from the E and CB compositions, assuming that it was incorporated as Si metal into the core. When the chondrite compositions are renormalized for this reduction in $SiO_2$, the resulting silicate mantle has lower $SiO_2$ and higher MgO contents, much closer to those in the estimates of Mercury's mantle obtained from the surface lavas (Table 2.2). The compositions of these $SiO_2$-reduced chondrites and the estimates of the mantle source regions for the NSP and IcP-HCT lavas are compared in Figure 2.6. There is a remarkable similarity between the silicate residues and the estimated Mercury mantle compositions. The largest discrepancies are in the high CaO of the IcP-HCT mantle and the high $Na_2O$ of the NSP mantle.

<Insert Figure 2.6 near here>

**2.5 Mercury's core composition**

As has been known since the middle of the last century, Mercury's high density indicates the presence of a central metallic core that makes up a substantially larger fraction of the body than do the cores of the other terrestrial planets (Anderson et al., 1987). MESSENGER-derived estimates of Mercury's moments of inertia have been used to refine that value to a core mass fraction of ~69–77 % (Hauck et al., 2013). Thus, Mercury's core has a substantial contribution to the bulk composition of the planet. However, determining the core composition of any planet is necessarily based on models, given that direct measurements are not possible. Even estimates of the composition of the best-characterized central metallic core in the solar system, the core of Earth, rely on models (e.g., Hillgren et al., 2000; McDonough, 2003; Li and Fei, 2014). One of the major constraints on these models comes from the fact that planetary cores make up a large fraction of their host planet, indicating that they must be composed of elements that are sufficiently abundant in the solar system to produce large cores. Nucleosynthesis processes result in Fe having much higher cosmochemical abundances than other neighboring elements, and considerable seismic and laboratory evidence supports Fe as the dominant component of Earth's core (e.g., Jeanloz, 1990; Li and Fei, 2014). Nickel is less abundant than Fe but is present in iron meteorites, some of which are likely samples of disrupted cores of differentiated planetesimals, at concentrations that are generally ~5–15 wt% and can range up to even 50 wt% (Goldstein et al., 2009). Metal in the highly reduced aubrite meteorites, a potentially relevant analog given Mercury's reduced nature (Burbine et al., 2002), contains an average of 5 wt% Ni (McCoy and Bullock, 2017). A bulk compositional model of the Earth based on chondritic concentrations gives ~5 wt% Ni in Earth's core (McDonough, 2003), though given the similarities between Fe and Ni, more detailed constraints are limited (Li and Fei, 2014). By similar logic, Mercury's core is most likely composed of dominantly Fe with Ni perhaps at a level similar to its abundance in Earth, aubrites, or iron meteorites.

Mercury's core is not solid but rather at least the outer part is molten (Margot et al., 2007; Chapters 4 and 19), indicating on the basis of thermal models that Mercury's core is not pure Fe-Ni, since the temperatures reached within Mercury would have resulted in such a core having frozen solid by the present (Schubert et al., 1988). In contrast, phase relationships in Fe-Ni systems that also contain a "light-element" component, i.e., one or more elements lower in atomic mass than Fe and Ni, enable Fe-Ni alloys to remain as liquids to lower temperatures (e.g., Li and Fei, 2014). The observation that Mercury's outer core is presently molten indicates that it contains a light-element component, and Monte Carlo modeling has been used to show that a large range of light element concentrations in Mercury's core can be consistent with the planet's radius, bulk density, and moment of inertia parameters (Hauck et al., 2013; Chapter 4). Thus, the identity and amount of the light-element component in Mercury's core are not highly constrained from measurements of the planet's bulk geophysical parameters, but they are important for understanding the planet's bulk composition, internal structure, and evolution, as well as the origin and history of its magnetic field (see, e.g., Chapters, 3, 4, 10, and 17).

Seismic data indicate that Earth's core also contains a light element component in addition to Fe and Ni (e.g., Jeanloz, 1990). Potential major light elements that have been considered to account for the density of Earth's core are H, C, O, Si, and S (e.g., Hillgren et al., 2000), due to their cosmochemical abundances and their potential affinities to alloy with Fe-Ni. Additionally, iron meteorites contain sulfides and carbides (Buchwald, 1975), and iron meteorite trace element trends support the presence of S during crystallization (e.g., Chabot and Haack, 2006), suggesting that S and C were incorporated into asteroidal cores and should be considered as potential light element components in larger planetary cores. Even though the large majority of Earth's P is predicted to be in its core and iron meteorites contain phosphides, P has a lower abundance in the solar system, and only 0.2 wt% P is estimated to be in Earth's core (McDonough, 2003). Hence, P is not considered as a major light element component. Applying a similar approach to evaluate the light element component of Mercury's core, we consider here H, C, O, Si, and S.

Hydrogen is the lightest and most abundant element in the universe, and a small amount of H in the core by mass would result in a considerable amount of H when evaluated in atomic fractions. Current consideration of H in Earth's core involves incorporation of H into metal during core formation in a hydrous magma ocean or from reaction at Earth's core-mantle boundary (e.g., Fukai, 1984; Okuchi, 1997; Williams and Hemley, 2001; Terasaki et al., 2012). However, the pressure, temperature, and oxygen fugacity conditions discussed for the incorporation of H into Earth's core are much higher than could be experienced by Mercury currently, given its reduced nature and core-mantle boundary pressure of ~5.5 GPa (Hauck et al., 2013). On the other hand, one hypothesis to explain Mercury's large core fraction involves an impact between bodies comparable or larger in mass than Mercury (see Chapter 18); we discuss the potential implications of such a scenario in more detail after evaluating other potential light element core constituents.

The measurement of ~1–5 wt% C associated with specific geologic units on Mercury's surface (Section 2.3.3) indicates that C was likely available in Mercury's interior during the formation of Mercury's core. Carbon alloys strongly with Fe under a wide range of conditions, and experiments on metal-silicate mixtures with application to early Earth show that more C partitions into the metallic phase than into the silicate phase by a factor of ~200–5000 (Dasgupta et al., 2013; Chi et al., 2014). Thus, the identification of C within the silicate portion of Mercury could imply that Mercury's core contains a substantial amount of C (up to a few wt%, Lord et al., 2009). However, recent work (Li et al., 2015) has also shown that C partitions less strongly into the metallic phase at reducing conditions (log $fO_2$=IW-4.7) and for Si-bearing metals (10 wt%). Such experiments still contain ~200 times more C in the metallic phase than the silicate phase, but preliminary experiments at even more reducing conditions with higher Si contents in the metal suggest that the solubility of C in metal may decrease further (Li et al., 2016; Vander Kaaden et al., 2016), such that C may not be a major light-element component in Mercury's core if it formed under highly reducing conditions. One implication of scenarios with high amounts of C in both Mercury's silicate portion and metallic core is that it would imply a bulk composition for Mercury that is substantially more C-rich than the other terrestrial planets. Alternatively, the high levels of C measured on Mercury's surface may indicate that most C did not partition into Mercury's core and that Mercury experienced different conditions during core formation than those of Earth, perhaps extremely reducing. The proposed hypothesis of graphite as a stable phase in Mercury's early history (Vander Kaaden and McCubbin, 2015) could sequester much, but not all, of Mercury's C budget if conditions were such that Mercury's core was C saturated and graphite was stable during Mercury's core formation.

Oxygen is an abundant element in the solar system and is a key component of the silicate portions of all of the terrestrial planets. At elevated temperatures, the solubility of O in metal increases. The high pressure and high temperature conditions of an early magma ocean on Earth could result in ~2–8 wt% O in Earth's core and the subsequent reduction of Earth's mantle (Rubie et al., 2004, 2015; Siebert et al., 2013; Tsuno et al., 2013; Fischer et al., 2015). In contrast, Mars is thought to not have experienced sufficiently high temperatures during its core formation to result in O in the martian core (Rubie et al., 2004), though reactions at the core-mantle boundary have been proposed to produce ~3 wt% O in the core of Mars (Tsuno et al., 2011). Thus, a body of Mercury's current size would not be predicted to result in any significant amount of O in the core.

Silicon is also an abundant and key element in the rocky portions of terrestrial planets, though under reducing conditions, it also alloys strongly with Fe metal (Kilburn and Wood, 1997; Hillgren et al., 2000; Gessmann et al., 2001; Malavergne et al., 2004). Thus, the reduced nature of Mercury's surface, as indicated by its low apparent FeO content (Section 2.1), led to the suggestion that Si is a major component of Mercury's core even prior to orbital results from MESSENGER (Malavergne et al., 2010), and the surface compositional results discussed above in Section 2.3, especially the high S abundance, provided further evidence of Mercury's reduced nature and the possibility of abundant Si in its core (Hauck et al., 2013; Chabot et al., 2014b; Malavergne et al., 2014; Namur et al., 2016b). Specific predictions of the amount of Si in Mercury's core are assessed in more detail below.

As discussed in Section 2.3.3, the discovery that Mercury's surface has several wt% S led to estimates of the highly reduced nature of the planet (oxygen fugacity of IW-3 to IW-7; McCubbin et al., 2012; Zolotov et al., 2013), since the solubility of S in silicate is known to increase with decreasing oxygen fugacity (e.g., Haughton et al., 1974; Berthet et al., 2009). For log $f$O$_2$ levels lower than approximately IW-4, more S partitions into the silicate phase than the metallic phase (Kilburn and Wood, 1997; Berthet et al., 2009), though the precise partitioning behavior is also dependent on the pressure, temperature, and composition conditions (Boujibar et al., 2014; Namur et al., 2016b). Thus, the presence of high levels of S on Mercury's surface could indicate that Mercury's core formed under reducing conditions and that S is not a major component of the core. Considerations of the cosmochemical abundance and volatility of S led Dreibus and Palme (1996) to conclude that even though the majority of Earth's S is believed to be in the core, Earth's core cannot contain more than 1.7 wt% S. Consequently, if Mercury's silicate fraction and its core both contain wt% levels of S, the bulk composition of Mercury could be more S-rich than Earth.

Having introduced each potential light element, we now return to a consideration of the implications for Mercury's core composition if the modern planet is only a smaller remnant of a once larger body, such as predicted if Mercury's large core is a result of a large impact event. Proposed impact scenarios include a proto-Mercury that was a factor of 2–5 more massive than the present planet involved in an impact with either a body ~20–40% the mass of current Mercury (Benz et al., 2007) or one much larger, at 85% the mass of Earth, in a hit-and-run collision (Asphaug and Reufer, 2014). All of these bodies considered are sufficiently large to have differentiated and formed central metallic cores prior to being involved in an impact event; the compositions of such cores are highly unconstrained and may have formed within bodies ranging in size from 20% of the mass of Mercury to 85% of the mass of Earth. During and following the impact event, the simulations of both Benz et al. (2007) and Asphaug and Reufer (2014) seem to show some level of mixing between the two cores to create Mercury's final core. The large-scale chemical evolution that would follow such a large impact event has not been examined. The energetics suggest that complete melting of the planet would have been possible, but the level of equilibration that would occur between a central metallic core largely present prior to the giant impact and the small fraction of silicate remaining is not clear, even if a global magma ocean was formed. Thus, if Mercury's large core is the result of a giant impact event, it may be even more challenging to constrain Mercury's core composition from measurements of its present silicate surface.

Though our current knowledge of Mercury cannot enable a unique identification of the core's composition, investigations that model the geophysical and geochemical evolution of the planet can place valuable constraints on abundances of light elements. From early MESSENGER measurements of Mercury's gravity field together with Earth-based measurements of Mercury's spin axis position and physical libration amplitude, Smith et al. (2012) suggested that a solid layer of FeS might be located at the top of the core, overlying a S- and Si-bearing fluid core. Subsequent revision to Mercury's obliquity (Margot et al., 2012) and modeling by Hauck et al. (2013) of Mercury's internal structure showed that a solid FeS layer was consistent with but not required by Mercury's geophysical parameters. Hauck et al. (2013) found numerous core composition solutions that ranged from 0–20 wt% S to 0–17 wt% Si and combinations of the two elements; extremely S-rich core compositions of >25wt% were also examined but were not favored because of the high S/Fe ratio that would be implied for the planet in comparison with all other known solar system materials. Overall, Hauck et al. (2013) concluded that Mercury's core likely contains a substantial quantity of Si and/or S.

Taking a geochemical approach, Malavergne et al. (2010) examined the core composition that would result from the differentiation of a reduced, metal-rich (EH or CB, see Section 2.4) chondrite bulk composition. They concluded that Mercury's core likely contains at least 5 wt% Si and is composed of at least two distinct liquids, one Fe-Si with almost no S and another Fe-S with almost no Si. Following up on this work with the incorporation of MESSENGER's surface composition measurements, Malavergne et al. (2014) concluded that sulfides likely played a major role in Mercury's differentiation, both by creating a sulfide layer at the top of the core and by enabling sulfides to be stable within the silicate portion of the planet.

In an alternative geochemical approach, Chabot et al. (2014b) performed metal-silicate experiments to examine the combined partitioning behavior of both S and Si under a range of oxygen fugacity conditions. They found that metallic melts with a range of S and Si contents could be in equilibrium with a silicate melt with wt% levels of S, as measured by MESSENGER for Mercury's surface. The experiments illustrated that constraints could be placed on the specific combinations of S and Si that were consistent with equilibrium core formation in a magma ocean scenario. They concluded that Mercury's core likely contains Si, limiting the S content of the core to <20 wt%, and for core Si contents >10 wt% they predicted that < 2 wt% S is also in the core.

Most recently, Namur et al. (2016b) experimentally examined the solubility of sulfur in reduced silicate melts, using compositions and oxygen fugacity conditions of relevance to Mercury on the basis of MESSENGER results. Combining their extensive experimental results with those of previous studies, they developed a parameterization expression to model the S concentrations in Mercury's magmas, mantle, and core. They concluded that Mercury's core has <1.5 wt% S and is Si-bearing, and that a thin, <90-km thick, layer of either molten or solid FeS-rich material may be present at the core-mantle boundary, depending on Mercury's bulk S content.

Other potential constraints on the composition of Mercury's core may be derived from modeling the thermal evolution and associated global contraction of the planet (see also Chapter 10). The abundance of large scarps and other contractional geologic features on Mercury's surface as measured from MESSENGER data indicates that the planet's radius has decreased by as much as 7 km since the late heavy bombardment (Byrne et al., 2014). Pre-MESSENGER models, based on a smaller estimate for the total global contraction and considering only S as a potential light element in the core, were used to make estimates of the core's composition and indicated >~6 wt% S (Hauck et al., 2004; Grott et al., 2011). Similar models that include the possibility of Si in the core and account for Mercury's larger measured contraction have the potential to provide new constraints on Mercury's core composition.

Similarly, the presence of a dynamo-driven modern magnetic field (Ness et al., 1975; Anderson et al., 2011; Chapter 5) as well as evidence from crustal remanent magnetism for an ancient magnetic field (Johnson et al., 2015) also potentially provide constraints on the composition of Mercury's core. Previous studies have often considered only S as a candidate light element in Mercury's core (Riner et al., 2008; Rivoldini et al., 2009; Manglik et al., 2010; Vilim et al., 2010; Dumberry and Rivoldini, 2015), but future models that also incorporate Si would be worthwhile, especially given the much higher solubility of Si in solid Fe metal (Kuwayama and Hirose, 2004) and potential challenges for compositional convection in the core to drive a dynamo.

Overall, Mercury's molten outer core indicates the presence of a light-element component, but the identity and abundance of that light-element component is poorly constrained. Table 2.3 summarizes the results of recent studies and suggests a growing consensus since MESSENGER's measurements that Mercury's core is likely Si-bearing and perhaps contains S and C as well.

The light-element composition of Mercury's core and the conditions during Mercury's core formation would also influence the concentrations of minor and trace elements in Mercury's core. In particular, understanding the concentrations of the heat-producing elements of K, U, and Th in Mercury's core is important, as their distribution has implications for the overall thermal evolution of the planet (Chapter 17). On the basis of Mercury's reduced nature and experimental metal-silicate partitioning data, Malavergne et al. (2010) concluded that U could be an important heat-producing element in Mercury's core whereas Th and K would not enter Mercury's core in significant amounts. However, in addition to oxygen fugacity, the S content of the phase can have a strong influence on the partitioning behavior of these elements, and if Mercury's core formed an FeS layer, such a layer could contain high concentrations of these heat-producing elements (McCubbin et al., 2012). If significant amounts of K, U, or Th are sequestered in Mercury's core, there also would be implications for interpreting the ratios of these elements measured by MESSENGER at Mercury's surface. Lastly, sulfides formed under reducing conditions in enstatite chondrite melting experiments contained elements that are more commonly lithophile, such as Mg, Ca, Cr, Mn, and Ti (McCoy et al., 1999; Berthet et al., 2009). Depending on the oxygen fugacity and partitioning of S during Mercury's core formation, Mercury's core could similarly contain minor amounts of elements not generally considered to be in planetary cores under more oxidizing conditions.

## 2.6 Mercury's bulk composition

Here we use the estimated mantle and core compositions discussed in the previous sections to estimate the bulk composition of Mercury for key elements. In section 2.4 we used petrologic results to estimate the compositions of mantle sources of two different measured surface compositions (Table 2.2). Since these two mantle estimates represent original compositions prior to crustal extraction, we take their average to be the bulk composition of Mercury's silicate shell, but we add 1 wt% Fe to match the surface abundance and 7 wt% S to match the mantle prediction of Namur et al. (2016b). From the geophysical modeling results of Hauck et al. (2013), we take Mercury's core radius to be 2020 km and its density to be 6980 kg m$^{-3}$, the silicate shell's density to be 3380 kg m$^{-3}$, and the planet's radius to be 2440 km. As discussed in detail in the previous section, Mercury's core likely contains Si, and possibly S and C, but the precise composition is highly uncertain (Table 2.3), so we consider a range of possibilities. For simplicity, we assume that the core has a fixed abundance of 1.5 wt% S (Namur et al., 2016b), and we vary the Si abundance from 0 to 25 wt%. We also consider cases with and without a 100-km-thick layer of FeS at the base of the mantle.

Selected element ratios from the resulting bulk compositions are compared in Figure 2.7 to the measured surface and inferred original mantle ("Bulk Sil.") compositions and the compositions of E and CB chondrites, Earth, and the Sun. The long-recognized effect of Mercury's large core is apparent: Mercury's bulk composition is highly Fe-rich compared with the bulk solar composition and that of other terrestrial planets and chondrites. However, Mercury's high iron fraction has often been expressed as the planet having a high Fe/Si ratio. The discovery from MESSENGER data that Mercury is extremely chemically reduced and consequently may have substantial Si in its core shows that, in fact, bulk Mercury is not only enriched in Fe, but also in Si and possibly S, relative to other planets, and in principle could have a chondritic Fe/Si ratio, but very low ratios of other elements to Si (see, e.g., the 25 wt% core Si cases in Figure 2.7). These plots reinforce the decades-long view that Mercury's origin must entail some form of metal-silicate fractionation, though whether this occurred as a result of chaotic (e.g., giant impacts) or orderly (e.g., nebular metal-silicate fractionation) processes is still unknown and will require substantial improvements in quantitative modeling of the chemical consequences of various proposed formation scenarios (see Chapter 18).

<Insert Figure 2.7 near here>

As discussed in Section 2.4, however, the estimated primitive mantle compositions are remarkably similar to the silicate compositions of E and CB chondrites if some Si is assumed to have been reduced to metallic form in the meteorite compositions. Indeed, for elements that don't partition into the core, Mercury's bulk composition is much closer to those of the other terrestrial planets, for example, Mercury's chondritic-like Al/Mg ratio (Figure 2.7c) and Mars-like K/Th and K/Cl ratios (Figure 2.3). The presence at chondritic levels of elements such as Al, K, and Na that preferentially partition into volcanic melts (e.g., surface Al/Mg ratio, Figure 2.6c) may provide constraints on Mercury origin scenarios. As discussed in Chapter 18, two types of impact models that have been discussed for Mercury's origin are a single giant impact stripping much of the silicate shell of a larger planet (e.g., Benz et al., 2007) and a hit-and-run scenario whereby a proto-Mercury impacts another body and transfers much of its silicate shell to the other planet before being scattered to the inner edge of the protoplanetary disk (e.g., Asphaug and Reufer, 2014). In the first scenario, the disrupted silicate material may be very well mixed before some of it re-accretes to the planet and thus may preserve the initial bulk volatile-rich composition (Section 18.4). However, if a proto-Mercury has already experienced a magma ocean stage and subsequent crystallization before being involved in a hit-and-run impact, incompatible elements may be enriched in the outer layers, and these may be preferentially lost. Mercury's present bulk silicate composition may thus argue against such an origin, but additional quantitative modeling is clearly needed.

<Insert Figure 2.8 near here>

The average of the Mercury mantle composition estimates derived in Section 2.4 is compared with the estimated compositions of the mantles of Earth (Hart and Zindler, 1986) and Mars (Dreibus and Wänke, 1985) on a plot of composition versus heliocentric distance in Figure 2.8. With the caveat that these mantle estimates are highly uncertain, especially for minor elements such as Ti, some characteristics of the silicate mantles are consistent with condensation temperatures that existed in the parts of the early solar nebula from which the bulk of material that accreted to form each planet originally condensed. There is a continual decrease in the mantle abundances of the refractory elements CaO, $Al_2O_3$, and $TiO_2$ from Mercury to Earth to Mars, which is consistent with Mercury having condensed in the hotter part of the solar nebula where CaO-, $Al_2O_3$-, and $TiO_2$- bearing minerals would have initially condensed. The continual increase in oxidized Fe from near zero for Mercury to a maximum of 17.5 wt% in the mantle of Mars is consistent with decreasing condensation temperatures and also with increasing $fO_2$ of the condensed minerals. However, other characteristics are much more enigmatic. Clearly there is an apparent decreasing compositional gradient in $SiO_2$ with distance from the Sun, with Si-rich enstatite chondrite-like material most abundant in Mercury's mantle. Neither this nor the observation that Mercury is the most volatile-rich planet (e.g., Na) would be expected from classical condensation sequences as one moves to lower temperatures away from the Sun (e.g., Grossman, 1972). However, the low oxygen fugacity inferred for Mercury must play a key role. Ebel and Alexander (2011) showed that condensation from a system enriched in C-rich, water-poor dust analogous to cometary dust particles can produce highly reduced, Fe-rich assemblages that could be precursors to E chondrites and Mercury. Moreover, under the inferred conditions, S, K, and Cl can all act as refractory elements (Ebel and Sack, 2013). Therefore, Mercury's volatile-rich and reduced silicate composition (and high overall Fe content in part) can be the result of condensation under highly reducing conditions perhaps influenced by exclusion of water ice from the feeding zone of Mercury's parental materials due to proximity to the Sun (Ebel and Alexander, 2011; Chapter 18).

## 2.7 Outlook

After decades of researchers arguing over theories in the face of a severe paucity of geochemical data for Mercury (Section 2.1), MESSENGER and its geochemical instruments provided the first survey of the innermost planet's surface composition for a wide range of elements, as detailed in Section 2.3. These data clearly revealed Mercury to be a volatile-rich but highly chemically reduced world, the implications of which for planet formation and evolutionary processes in our solar system and in exoplanetary systems are still poorly understood. Although the fundamental scientific question of why Mercury is so metal-rich is still unanswered, the importance of having data by which to test ideas and models cannot be overstated, and the MESSENGER compositional data set is rich indeed.

Scientific implications of the compositional data for Mercury's origin and geological history are discussed throughout this book (e.g., Chapters 7, 8, 10, 11, 12, 18, and 19), but the MESSENGER observations have opened a vast window for new questions and future work. Petrologic processes under highly reducing conditions are still not very well understood, but, largely motivated by MESSENGER results, are being addressed increasingly by experimental studies (McCoy et al., 1999; Berthet et al., 2009; Charlier et al., 2013; Chabot et al., 2014b; Namur et al., 2016b, 2016b; Vander Kaaden and McCubbin, 2016). Such studies are needed to fully understand the origin of Mercury's geochemical terranes, the nature of its surface rocks, melting conditions in the planet's mantle, and element partitioning into Mercury's core. The discovery of abundant C on Mercury's surface (Section 2.3.3) indicates the need for experimental and theoretical work that includes this element. As discussed in detail in Chapter 18, a great deal of further theoretical work is needed to assess the array of proposed origin scenarios to account for Mercury's high metal-to-silicate ratio and especially to evaluate chemical consequences. The MESSENGER geochemical data set will be crucial to testing these scenarios. Finally, the availability of element abundance maps on Mercury's surface will be useful for better understanding Mercury's exosphere and the surface processes that supply it.

Looking beyond the MESSENGER data, the BepiColombo mission (Chapter 20) will carry a suite of geochemical instruments similar to those of MESSENGER (X-ray, gamma-ray, and neutron spectrometers) and will provide complementary – and for some areas improved – geochemical information about Mercury. BepiColombo's polar orbit will be much less eccentric than that of MESSENGER and thus provide southern hemisphere elemental measurements that were either not possible before (i.e., GRS measurements) or were made with poor spatial resolution (XRS). BepiColombo's thermal imaging spectrometer may provide crucial mineralogical data, not obtained with MESSENGER instruments, which can be combined with elemental abundance data to improve our understanding of the geological and space-weathering history of the surface.

Table 2.1. Surface elemental composition of Mercury[a].

| Element (ratio) | XRS | GRS[b] | NS[b] |
|---|---|---|---|
| K (ppm, average) | | 1288 ± 234 | |
| K (ppm, range) | | 240–2500 | |
| Th (ppm) | | 0.155 ± 0.054 | |
| U (ppb) | | 90 ± 20 | |
| K/Th | | 8000 ± 3200 | |
| Mg/Si | 0.436 (0.106) | | |
| Al/Si | 0.268 (0.048) | $0.29^{+0.05}_{-0.13}$ | |
| S/Si | 0.076 (0.019) | 0.092 ± 0.015 | |
| Ca/Si | 0.165 (0.030) | 0.24 ± 0.05 | |
| Ti/Si | 0.012 ± 0.001 | | |
| Cr/Si | 0.006 ± 0.001 | | |
| Mn/Si | 0.004 ± 0.001 | | |
| Fe/Si | 0.053 (0.013) | 0.077 ± 0.013 | |
| Na/Si (average) | | 0.12 ± 0.01 | |
| Na/Si (0–60°N) | | 0.107 ± 0.008 | |
| Na/Si (80°–90°N) | | 0.198 ± 0.030 | |
| Cl/Si (average) | | 0.0057 ± 0.0010 | |
| Cl/Si (0–60°N) | | 0.0049 ± 0.001 | |
| Cl/Si (80°–90°N) | | 0.014 ± 0.005 | |
| O/Si | | 1.2 ± 0.1 | |
| C (wt%) | | 1.4 ± 0.9 | ~1–4 |

[a]Ratios are by mass. Numbers in parenthesis indicate the standard deviation (σ) of the XRS measurements, reflecting surface variability; ± symbol denotes the 1σ statistical uncertainty.

[b]GRS and NS data are from the northern hemisphere.

Data sources: Nittler et al. (2011, 2016); Peplowski et al. (2011, 2012b, 2014, 2015a, 2016); Evans et al. (2012, 2015); Weider et al. (2014, 2015); Frank et al. (2015); McCubbin et al. (2017)

Table 2.2 Estimates of Mercury mantle silicate composition[a]

|  | SiO$_2$ | TiO$_2$ | Al$_2$O$_3$ | FeO | MgO | CaO | Na$_2$O | K$_2$O |
|---|---|---|---|---|---|---|---|---|
| Mantle melts and residual minerals[b] | | | | | | | | |
| NSP lava | 58.7 | 0.4 | 13.8 | 0.04 | 13.9 | 5.81 | 7 | 0.2 |
| IcP-HCT lava | 52.7 | 0.4 | 8.79 | 0.04 | 27.8 | 7.27 | 2.75 | 0.08 |
| Enstatite, NSP residuum | 58.8 | 0.32 | 2.15 | 0.01 | 38.6 | 1.55 | 0.21 | 0 |
| Enstatite, IcP residuum | 59.3 | 0.07 | 0.65 | 0.01 | 37.2 | 1.44 | 0.07 | 0 |
| Forsterite residuum | 43.3 | | | 0.02 | 55.78 | 0.23 | | |
| Pre-melting mantle composition from mass balance | | | | | | | | |
| NSP | 53.67 | 0.24 | 4.57 | 0.02 | 36.89 | 2.26 | 1.97 | 0.05 |
| IcP-HCT | 51.98 | 0.21 | 4.24 | 0.03 | 37.64 | 3.84 | 1.29 | 0.04 |
| Chondrite compositions | | | | | | | | |
| Eagle (EL6)[c] | 60.6 | 0.14 | 2.96 | 0 | 33.57 | 1.05 | 1.16 | 0.13 |
| Eagle (EL6), Si reduced[d] | 55.16 | 0.16 | 3.37 | 0 | 38.2 | 1.2 | 1.33 | 0.14 |
| ALHA77295 (EH4)[c] | 61.21 | 0.19 | 3.07 | 0 | 31.45 | 2.16 | 1.34 | 0.12 |
| ALHA77295 (EH4). Si reduced[d] | 51.98 | 0.21 | 3.5 | 0 | 35.83 | 2.46 | 1.52 | 0.14 |
| Bencubbin (CB)[c] | 61.55 | 0.25 | 2.47 | 0 | 32.58 | 2 | 1.02 | 0.13 |
| Bencubbin (CB). Si reduced[d] | 56.15 | 0.29 | 2.81 | 0 | 37.16 | 2.28 | 1.17 | 0.15 |

[a] Not including the possible presence of 7–11 wt% sulfur (Namur et al., 2016b) or reduced Fe.

[b] From Namur et al. (2016a). NSP: northern smooth plains, IcP-HCT: intercrater plains and heavily cratered terrain; enstatite: MgSiO$_3$; forsterite: Mg$_2$SiO$_4$. Values in wt %.

[c] Silicate portion of analyses reported by Jarosewich (1990); Bencubbin is "Bencubbin II" analysis.

[d] Chondrite analyses from above, but with 20% of SiO$_2$ reduced to Si and assumed to have gone into the core.

**Table 2.3** Estimated major element abundances in Mercury's core.

| | Fe (wt%) | Ni (wt%) | C (wt%) | Si (wt%) | S (wt%) |
|---|---|---|---|---|---|
| Typical iron meteorites[1] | | ~5 - 15 | | | |
| Aubrite metal[2] | | 5 | | | |
| Earth model[3] | | 5 | | | |
| Model of graphite flotation crust[4] | | | >0 | | |
| Geophysical models[5] | ~75–95 | ~5[a] | | >0 - 17 | 0–20 |
| Metal-rich chondrite differentiation model[6] | ~75–89 | ~5[a] | | >5 | ~1–15 |
| Core formation experiments[7] | ~75–83 | ~5[a] | | >1 | 0–20 |
| S solubility experiments[8] | | | | >0 | <1.5 |

[a] Value assumed to estimate the abundance range of Fe

Data sources: 1. Goldstein et al. (2009); 2. McCoy and Bullock (2017); 3. McDonough (2003); 4. Vander Kaaden and McCubbin (2015); 5. Hauck et al. (2013); 6. Malavergne et al. (2010); 7. Chabot et al. (2014b); 8. Namur et al. (2016b)

**Figure Captions**

**Figure 2.1** (a) Example X-ray spectra from Mercury recorded by the three planet-facing detectors on MESSENGER's XRS. Spectra acquired during typical quiet-Sun conditions (solid curves) show fluorescence from Mg, Al, and Si, whereas spectra during solar flares (dashed curves) have much higher signal-to-background ratios and show fluorescence from elements up to Fe. Adapted from Weider et al. (2015). (b) Gamma-ray spectra measured by MESSENGER's GRS during the first two months of orbital operations. The high-altitude (>7500 km; red) spectrum samples instrument background, and the low-altitude (<1200 km; black) spectrum samples both background and signals from Mercury. The difference spectrum (blue) highlights the Mercury component. Several strong gamma-ray lines are labeled by element; the prominent line at 511 keV is due to electron-positron annihilation

**Figure 2.2** Elemental abundance and weight ratio maps for Mercury, derived from MESSENGER GRS (K; Peplowski et al. 2012a) and XRS (Weider et al. 2015; Nittler et al. 2016) data. Maps are shown in a Molleweide projection, centered on 0°N, 0°E. Red circles with one-standard-deviation errors are northern-hemisphere averages from GRS (Table 2.1); red lines in color scales are area-weighted global averages of mapped data. HMR indicates the location of the high-Mg region, and CB is the Caloris impact basin.

**Figure 2.3** (a) K/Th mass ratios for inner solar system bodies as a function of heliocentric distance. Values and uncertainties are derived from Peplowski et al. (2012a), Lodders and Fegley (1998), Prettyman et al. (2006), Taylor et al. (2006), and Prettyman et al. (2015). (b) Mercury's surface Cl and K concentrations compared with those of other solar system objects. Adapted from Evans et al. (2015).

**Figure 2.4** (a) Mg/Si versus Al/Si (elemental mass ratios) on Mercury compared with terrestrial and lunar rock compositions; after Nittler et al. (2011). Red contours indicate the distribution of Mercury values (from Figure 2.2), with the outer contour enclosing 99% of the data and other contours representing 5% increments. Blue stars indicate compositions modeled by Namur et al. (2016a); see Table 2.2. The black triangle is the composition of a 1425 °C (29%) partial melt of the Indarch enstatite chondrite (McCoy et al., 1999). (b) Ca/Si versus S/Si (elemental weight ratios) for several hundred solar-flare XRS analyses with one-standard-deviation errors. Ca and S are highly correlated; the orange dashed line shows the best-fit line, which has a slope of 1.8. Rach. PD indicates a measurement of the pyroclastic volcanic deposit northeast of the Rachmaninoff impact basin. This deposit falls clearly off the correlation line, indicating loss of S. Adapted from Weider et al. (2016).

**Fig. 2.5** Schematic pressure–temperature diagram for partial melting of Mercury's mantle during adiabatic decompression. The IcP-HCT mantle adiabat is markedly hotter and crosses the solidus at a greater depth and higher temperature than that for the NSP. This higher temperature leads to a greater extent of melting. The solidus is the locus of pressure and temperature at which the mantle begins to partially melt. Dashed lines show the temperature at increasing extents of partial melting in 10 % increments.

**Figure 2.6** Estimated compositions (blue circles) of mantle sources of Mercury surface lavas from the high-Mg intercrater plains and heavily cratered terrain (IcP-HCT) and northern smooth plains (NSP) compared with compositions of the silicate portions of two enstatite and one Bencubbin-type chondrite with 20% of the $SiO_2$ removed and assumed to have been reduced to Si and sequestered in the core. Data from Table 2.2.

**Figure 2.7** Bulk compositions of Mercury (open and grey-filled circles) derived from estimates of mantle and core compositions (Sections 2.4 and 2.5) and the physical parameters of Hauck et al. (2013) compared with compositions of meteorites (Table 2.2, from Jarosewich, 1990), Mercury surface and bulk silicate (black circles, Sections 2.3 and 2.4), Earth, and Sun. Core compositions include 1.5 wt% S, and the mantle composition corresponds to the average of the two mantle estimates in Table 2.2 with 1 wt% Fe and 7 wt% S added. Different bulk compositions correspond to different assumed Si concentrations in Mercury's core and different assumptions regarding internal structure, as indicated.

**Figure 2.8** Comparison of Mercury's estimated mantle composition with those of Earth (Hart and Zindler, 1986) and Mars (Dreibus and Wänke, 1985), plotted versus heliocentric distance. Mercury compositions are averages of the two mantle estimates in Table 2.2 and Figure 2.6.

Head, J. W., Murchie, S. L., Prockter, L. M., Robinson, M. S., Solomon, S. C., Strom, R. G., Chapman, C. R., Watters, T. R., McClintock, W. E., Blewett, D. T. and Gillis-Davis, J. J. (2008). Volcanism on Mercury: Evidence from the first MESSENGER flyby. *Science,* **321**, 69–72.

Hillgren, V. J., Gessmann, C. K., Li, J., Righter, K., et al. (2000). An experimental perspective on the light element in Earth's core. In *Origin of the Earth and Moon*, ed. R. M. Canup and K. Righter. Tucson, AZ: The University of Arizona Press, pp. 245–263.

Jarosewich, E. (1990). Chemical analyses of meteorites: A compilation of stony and iron meteorite analyses. *Meteoritics,* **25**, 323–338.

Jeanloz, R. (1990). The nature of the Earth's core. *Ann. Rev. Earth Planet. Sci.,* **18**, 357–386. doi:doi:10.1146/annurev.ea.18.050190.002041.

Johnson, C. L., Phillips, R. J., Purucker, M. E., Anderson, B. J., Byrne, P. K., Denevi, B. W., Feinberg, J. M., Hauck, S. A., Head, J. W., Korth, H., James, P. B., Mazarico, E., Neumann, G. A., Philpott, L. C., Siegler, M. A., Tsyganenko, N. A. and Solomon, S. C. (2015). Low-altitude magnetic field measurements by MESSENGER reveal Mercury's ancient crustal field. *Science,* **348**, 892–895. doi:10.1126/science.aaa8720.

Kasper, R. B. (1975). Cation and oxygen diffusion in albite. Ph. D. thesis. Providence, RI: Brown University, 143 pp.

Kerber, L., Head, J. W., Solomon, S. C., Murchie, S. L., Blewett, D. T. and Wilson, L. (2009). Explosive volcanic eruptions on Mercury: Eruption conditions, magma volatile content, and implications for interior volatile abundances. *Earth Planet. Sci. Lett.,* **285**, 263–271.

Kilburn, M. R. and Wood, B. J. (1997). Metal-silicate partitioning and the incompatibility of S and Si during core formation. *Earth Planet. Sci. Lett.,* **152**, 139–148.

Kuwayama, Y. and Hirose, K. (2004). Phase relations in the system Fe-FeSi at 21 GPa. *Amer. Min.,* **89**, 273–276.

Lawrence, D. J., Feldman, W. C., Goldsten, J. O., McCoy, T. J., Blewett, D. T., Boynton, W. V., Evans, L. G., Nittler, L. R., Rhodes, E. A. and Solomon, S. C. (2010). Identification and measurement of neutron-absorbing elements on Mercury's surface. *Icarus,* **209**, 195–209.

Lawrence, D. J., Feldman, W. C., Goldsten, J. O., Maurice, S., Peplowski, P. N., Anderson, B. J., Bazell, D., McNutt, R. L., Nittler, L. R., Prettyman, T. H., Rodgers, D. J., Solomon, S. C.
41

.

Figure 2.1

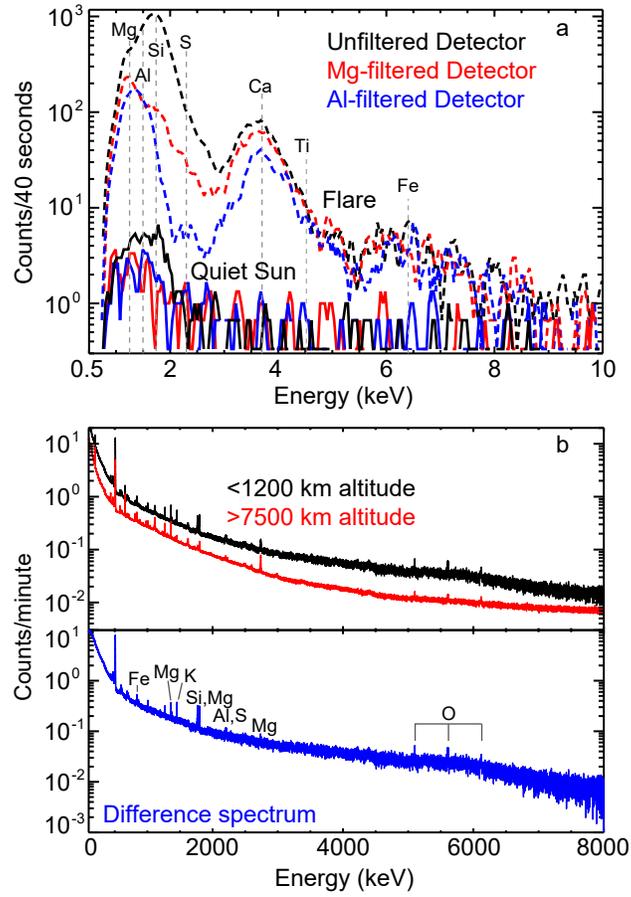

Figure 2.2

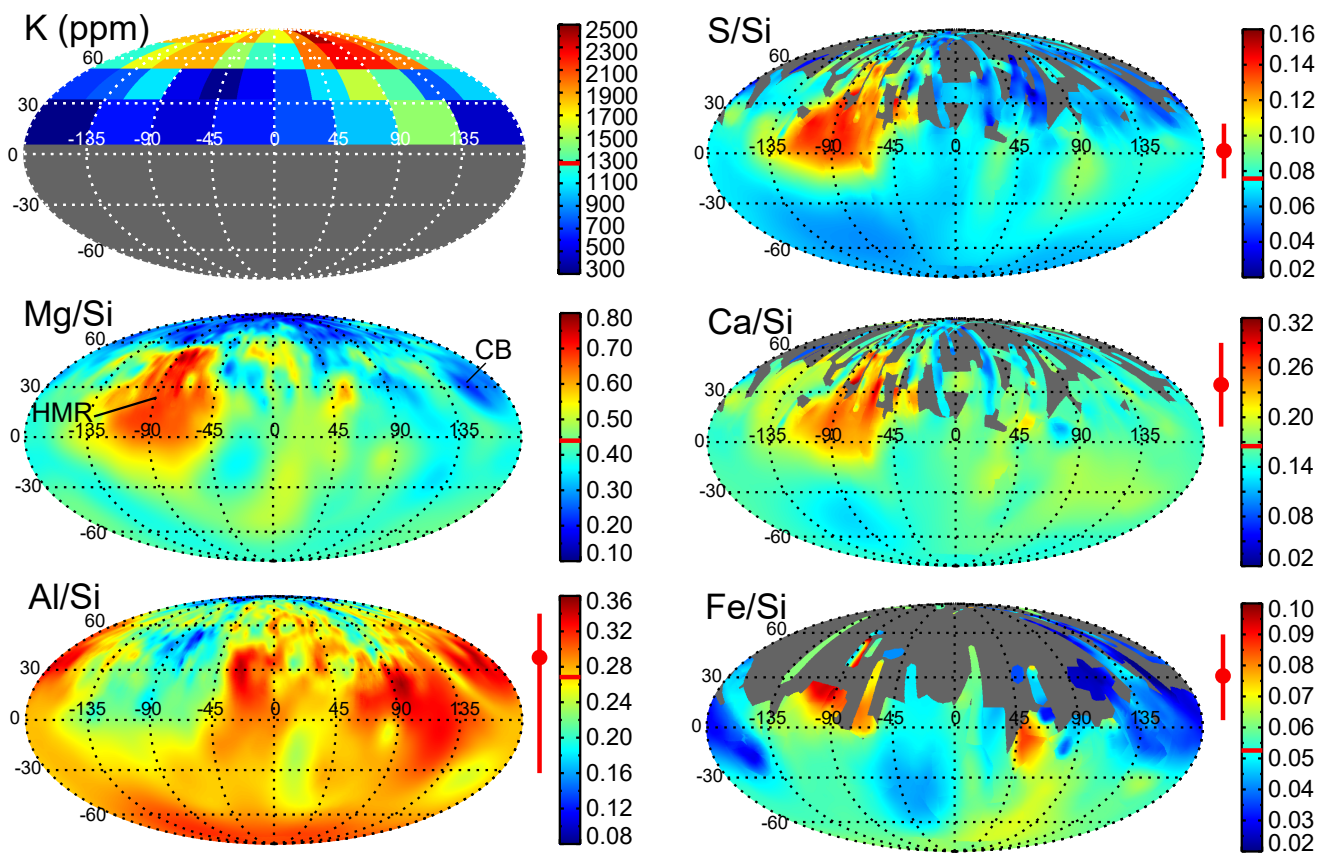

Figure 2.3

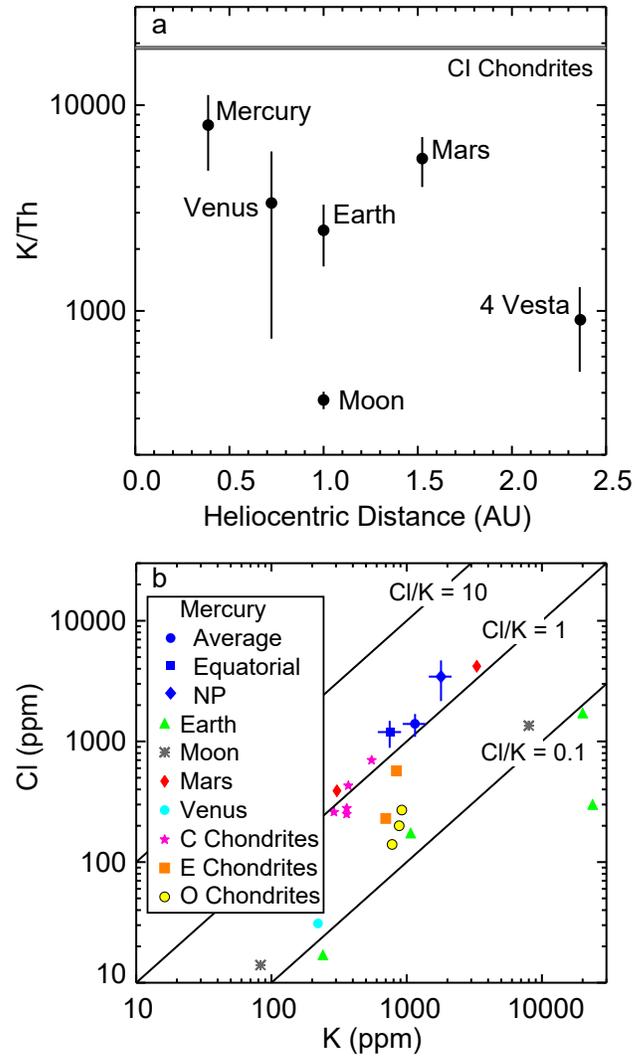

Figure 2.4

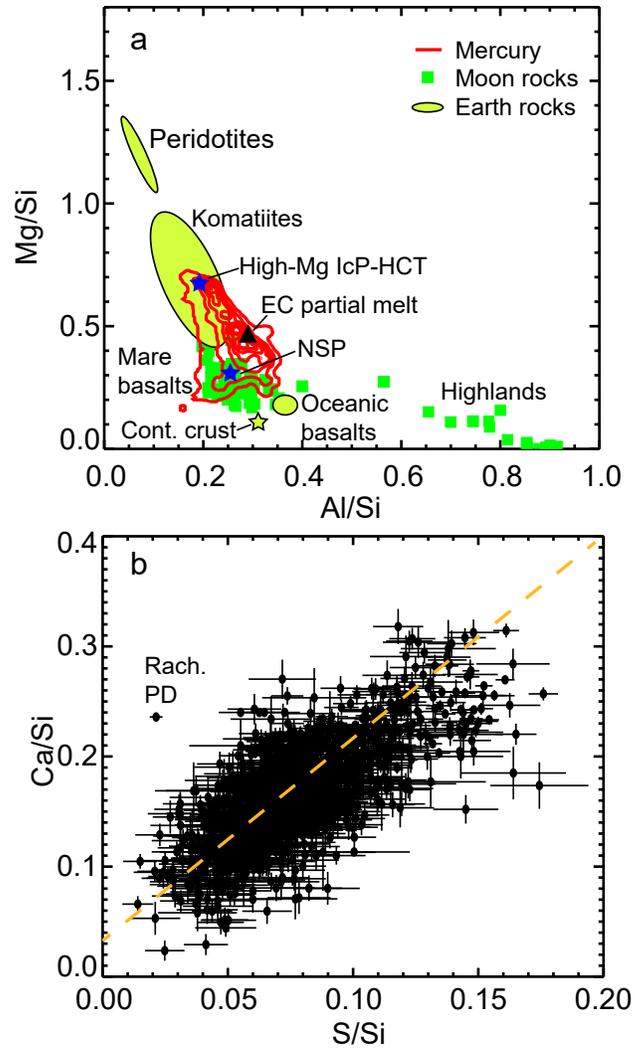

Figure 2.5

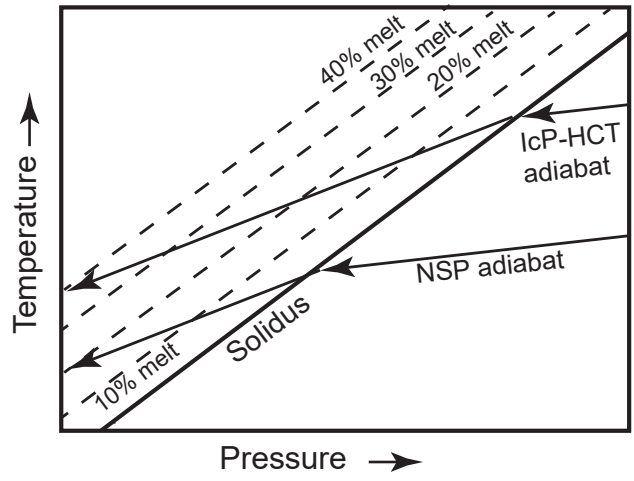

Figure 2.6

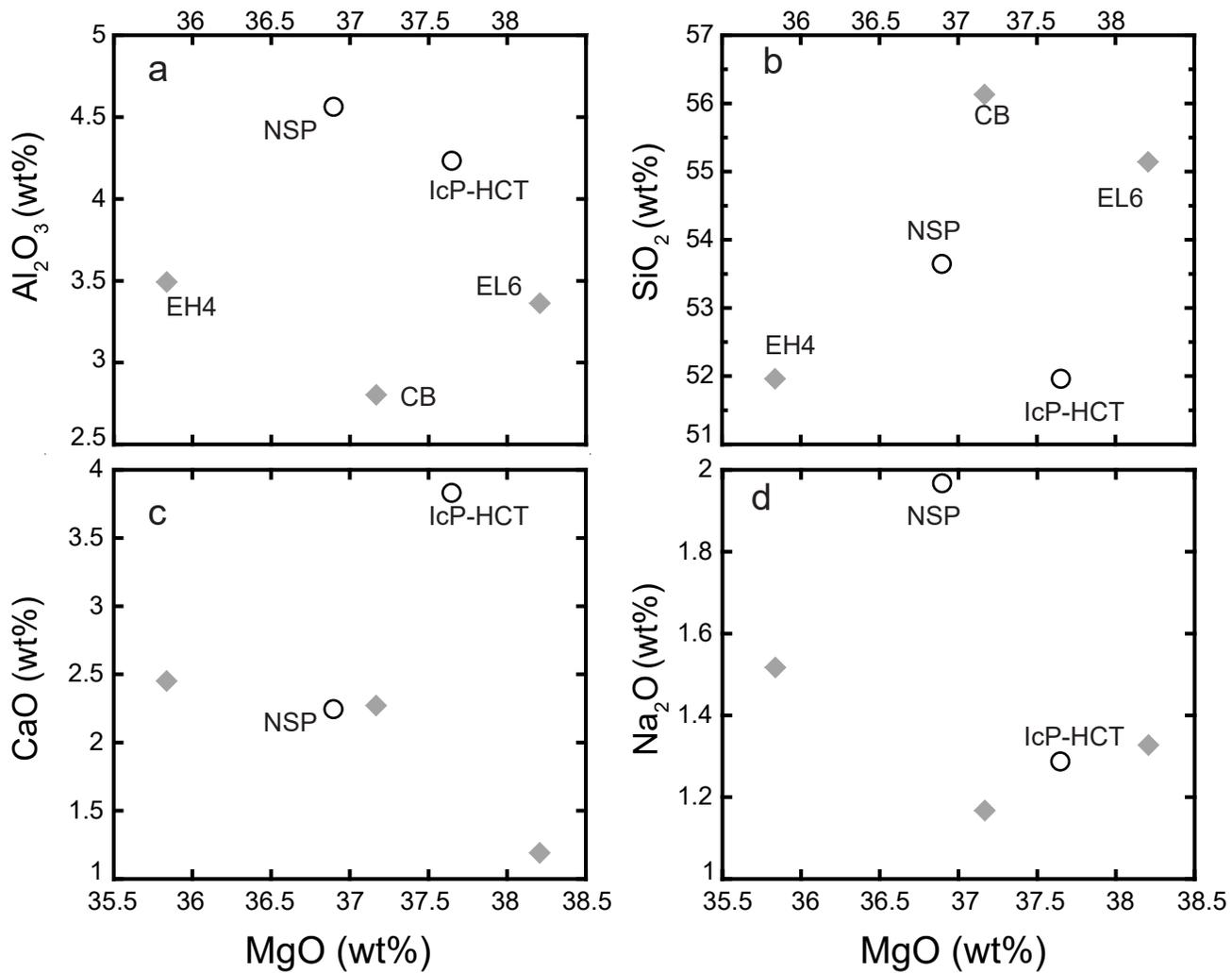

Figure 2.7

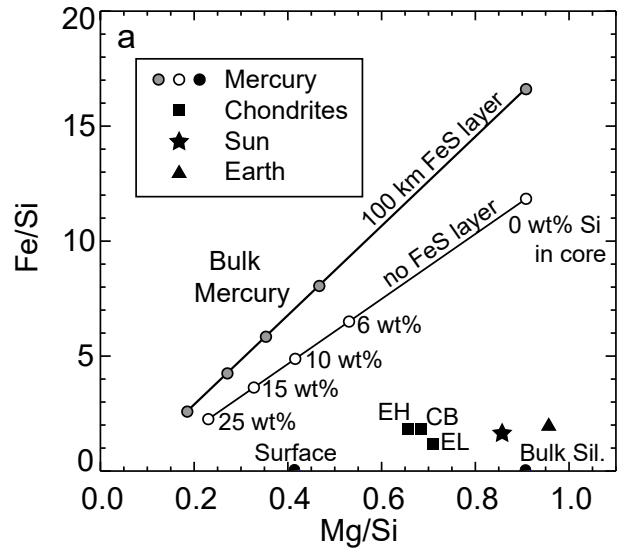

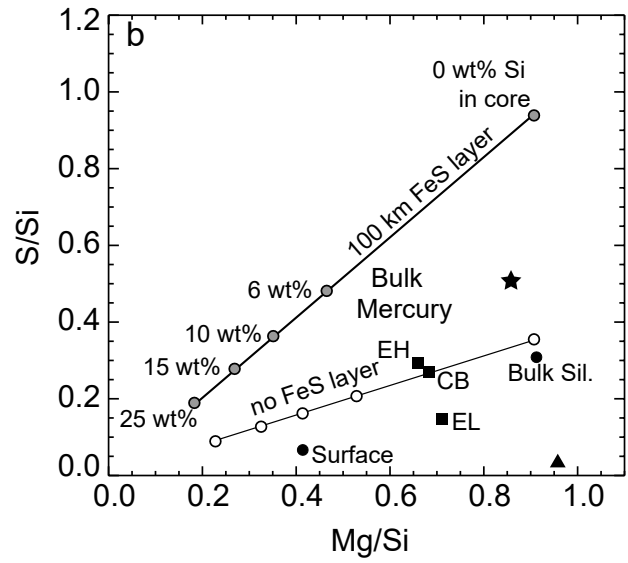

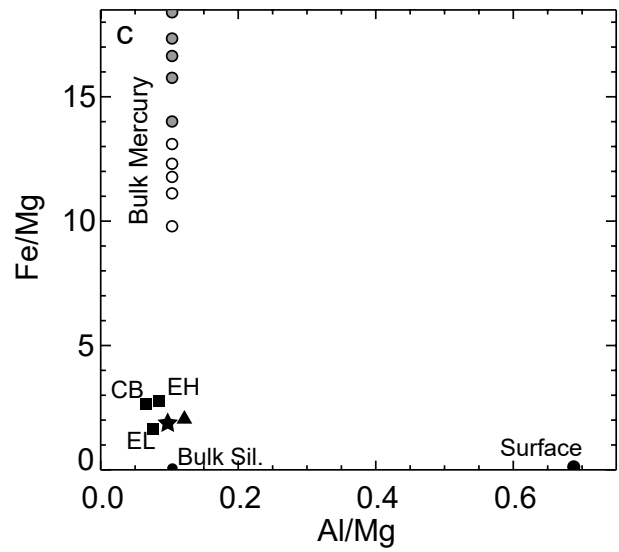

Figure 2.8

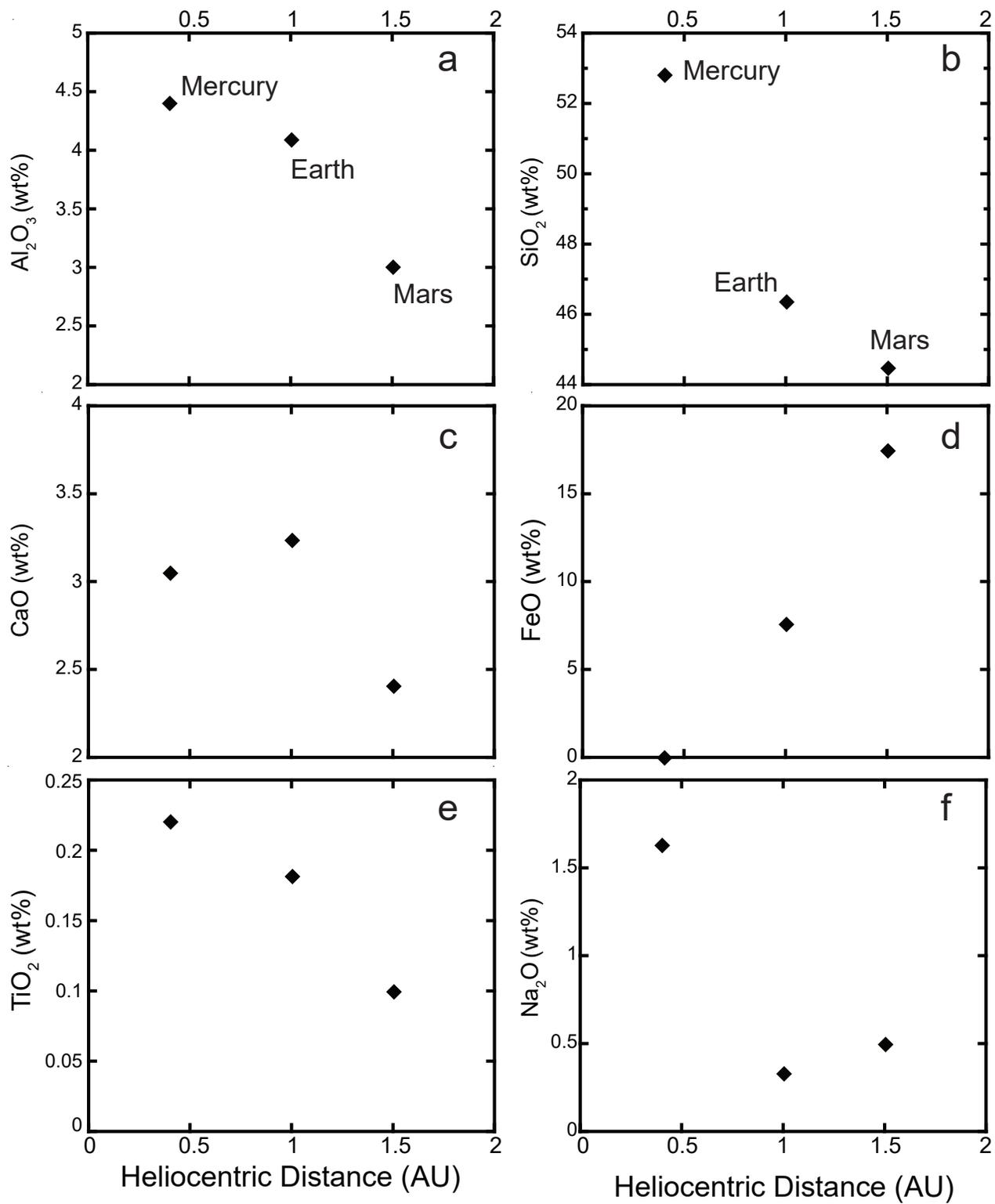